\documentclass{emulateapj}
\usepackage{amsmath}
\usepackage{txfonts}
\usepackage{color}
\usepackage{fancybox}
\newcommand{\unitspace}{\ensuremath{\,}}
\newcommand{\usp}{\unitspace}
\newcommand{\numberspace}{\ensuremath{\;}}
\newcommand{\nsp}{\numberspace}
\newcommand{\unitstyle}[1]{\ensuremath{\text{#1}}}
\newcommand{\power}[2]{\ensuremath{{#1}^{#2}}}

\newcommand{\nano}{\unitstyle{n}}

\newcommand{\centi}{\unitstyle{c}}
\newcommand{\kilo}{\unitstyle{k}}
\newcommand{\Mega}{\unitstyle{M}}
\newcommand{\Giga}{\unitstyle{G}}

\newcommand{\meter}{\unitstyle{m}}
\newcommand{\gram}{\unitstyle{g}}
\newcommand{\second}{\unitstyle{s}}

\newcommand{\cm}{\centi\meter}
\newcommand{\Kelvin}{\unitstyle{K}}
\newcommand{\K}{\Kelvin} 

\newcommand{\grampercc}{\gram\usp\power{\cm}{-3}} 

\newcommand{\eV}{\unitstyle{eV}}        
\newcommand{\keV}{\kilo\eV} 
\newcommand{\MeV}{\Mega\eV} 

\newcommand{\Msun}{\ensuremath{M_\odot}}

\newcommand{\yr}{\unitstyle{yr}}        
\newcommand{\km}{\kilo\meter}   

\newcommand{\kB}{\ensuremath{k_\mathrm{B}}} 
\newcommand{\NA}{\ensuremath{N_\mathrm{\!A}}} 

\newcommand{\ee}[1]{\ensuremath{\times 10^{#1}}}

\newcommand{\code}[1]{\textsc{#1}}

\newcommand{\nuclei}[2]{\ensuremath{\mathrm{^{#1}#2}}}

\newcommand{\neutron}{\ensuremath{n}}
\newcommand{\nt}{\neutron}
\newcommand{\proton}{\ensuremath{p}}
\newcommand{\pt}{\proton}
\newcommand{\hydrogen}[1][1]{\nuclei{#1}{H}}
\newcommand{\helium}[1][4]{\nuclei{#1}{He}}

\newcommand{\carbon}[1][12]{\nuclei{#1}{C}}
\newcommand{\nitrogen}[1][14]{\nuclei{#1}{N}}
\newcommand{\oxygen}[1][16]{\nuclei{#1}{O}}
\newcommand{\fluorine}[1][19]{\nuclei{#1}{F}}
\newcommand{\neon}[1][20]{\nuclei{#1}{Ne}}
\newcommand{\sodium}[1][23]{\nuclei{#1}{Na}}
\newcommand{\magnesium}[1][24]{\nuclei{#1}{Mg}}
\newcommand{\aluminum}[1][27]{\nuclei{#1}{Al}}

\newcommand{\iron}[1][56]{\nuclei{#1}{Fe}}

\newcommand{\nickel}[1][58]{\nuclei{#1}{Ni}}

\newcommand{\germanium}[1][74]{\nuclei{#1}{Ge}}

\newcommand{\SNeIa}{SNe~Ia}
\newcommand{\mean}[1]{\ensuremath{\langle #1 \rangle}}
\newcommand{\Abar}{\mean{A}}

\newcommand{\Sdot}{\ensuremath{\varepsilon}}
\newcommand{\Ye}{\ensuremath{Y_{e}}}
\newcommand{\tH}{\ensuremath{t_{\mathrm{H}}}}
\newcommand{\Fn}{\ensuremath{\tilde{F}}}

\newcommand{\qeff}{\ensuremath{q_{\mathrm{eff}}}}
\newcommand{\vrms}{\ensuremath{v_{\mathrm{rms}}}}

\begin{document}
\title{The Reduction of the Electron Abundance during the Pre-explosion Simmering in White Dwarf Supernovae}
\shorttitle{Pre-explosion simmering in Type Ia Supernovae}
\author{David A. Chamulak\altaffilmark{1,2}, Edward F. Brown\altaffilmark{1,2,3},
F. X. Timmes\altaffilmark{2,4,5}, and
Kimberly Dupczak\altaffilmark{1}}
\altaffiltext{1}{Department of Physics and Astronomy, Michigan State University, East Lansing, MI 48824}
\altaffiltext{2}{Joint Institute for Nuclear Astrophysics}
\altaffiltext{3}{National Superconducting Cyclotron Laboratory}
\altaffiltext{4}{Thermonuclear Applications, X-2, Los Alamos National Laboratory}
\altaffiltext{5}{School of Earth and Space Exporation, Arizona State University, Tempe, AZ 85287}
\email{chamulak@pa.msu.edu}
\email{ebrown@pa.msu.edu}
\email{fxt44@mac.com}
\email{dupczakk@msu.edu}
\shortauthors{Chamulak et al.}

\begin{abstract}
Prior to the explosion of a carbon-oxygen white dwarf in a Type Ia
supernova there is a long ``simmering,'' during which the
$\carbon+\carbon$ reaction gradually heats the white dwarf on a long
($\sim 10^{3}\nsp\yr$) timescale.  Piro \& Bildsten showed that weak reactions during this simmering set a maximum
electron abundance \Ye\ at the time of the explosion. We investigate
the nuclear reactions during this simmering with a series of
self-heating, at constant pressure, reaction network calculations.
Unlike in AGB stars, proton captures onto \neon[22] and
heavier trace nuclei do not play a significant role. 
The \carbon\ abundance is sufficiently high that the neutrons
preferentially capture onto \carbon, rather than iron group nuclei. 
As an aid to hydrodynamical simulations of the simmering phase, we
present fits to the rates of heating, electron capture, change in mean
atomic mass, and consumption of \carbon\ in terms of the screened
thermally averaged cross section for $\carbon+\carbon$. Our evaluation
of the net heating rate includes contributions from electron captures
into the 3.68\nsp\MeV\ excited state of \carbon[13]. This results in a slightly larger energy release, per \carbon\ consumed, than that found by Piro \& Bildsten, but less than that released for a burn to only \neon\ and \sodium. We compare our one-zone
results to more accurate integrations over the white dwarf structure
to estimate the amount of \carbon\ that must be consumed to raise the
white dwarf temperature, and hence to determine the net reduction of
\Ye\ during simmering. 

\end{abstract}


\section{Introduction}\label{s.introduction}

The leading scenario for a Type Ia supernova (hereafter \SNeIa) is the thermonuclear incineration of a carbon-oxygen white dwarf that has increased in mass, through accretion, to just below the Chandrasekhar limit \citep[for a review, see][]{hillebrandt.niemeyer:type}. Despite the advances in modeling the post-ignition flame evolution \citep[for a sampling of recent work, see][]{gamezo.khokhlov.ea:deflagrations,plewa.calder.ea:type,Ropke2005Type-Ia-superno,Jordan2007Three-Dimension}, we still lack a complete understanding of which subset of the binary white dwarf population become \SNeIa, and how differences in the progenitor map onto differences in the outcome of the explosion.

The composition of the white dwarf at the time of the explosion should
have an effect on the nucleosynthesis that takes place during the
explosion and the isotopic abundances of the final composition.  The
observable properties of \SNeIa\ resulting from Chandrasekhar-mass
explosions are chiefly determined by their final composition, the
velocity profiles of key spectral lines at early- and late-times
(e.g., P-Cygni profiles of \ion{Si}{2} at 615.0\nsp\nano\meter\ at
early times), the opacity of the material through which the photons
from radioactive decay must propagate, the kinetic energy of
ejecta, and its interaction with the density profile of the surrounding
circumstellar or interstellar medium \citep{filippenko:optical,pinto.eastman:physics,hillebrandt.niemeyer:type,leibundgut:cosmological,Mazzali2006The-54Fe58Ni/56,Marion2006Low-Carbon-Abun,Blondin2006Using-Line-Prof,Badenes2007Are-the-Models-,Woosley2006Type-Ia-Superno}.  The dominant parameter in setting the
peak brightness, and hence width, of the light curve is widely believed
to be the mass of \nickel[56] ejected by the explosion.
\citet{timmes.brown.ea:variations} showed the mass of \nickel[56]
produced depends linearly on the electron fraction, \Ye, at the time
of the explosion, and that \Ye\ itself depends linearly on the
abundance of \neon[22] in the white dwarf.

In this paper we explore, using a reaction network coupled to an
equation for self-heating at constant pressure, the reduction in \Ye\
that occurs after the onset of the thermonuclear runaway (when the
heating from $\carbon+\carbon$ reactions is faster than cooling by
thermal neutrino emission), but before the burning becomes so fast
that local regions can thermally run away and launch a flame. This
``simmering'' epoch lasts for $\sim 10^{3}\nsp\yr$, long enough that
electron captures onto products of \carbon\ burning can reduce the
free electron abundance \Ye. A similar, but independent calculation, of the
reduction of \Ye\ during simmering was performed by
\citet{Piro2007Neutronization-}.  Their calculation did not use a full
reaction network, but it did take into account the change in energy of
the white dwarf due to the growth of the convective zone.  Our
calculation agrees with their findings, in particular that there is a
maximum \Ye\ at the time of the explosion, and that the reduction in
\Ye\ is linear in the amount of \carbon\ consumed prior to when the
rate of \carbon\ burning outpaces that of the weak reactions. This
paper expands on their work in three ways. First, by using a full
reaction network, we are able to quantify the role of \neon[22] and
trace nuclides in setting the change in electron abundance with
\carbon\ consumption, $d\Ye/dY_{12}$. Second, we calculate the heating
from the electron capture reactions and include the contribution from
an excited state of \carbon[13]. Third, we provide tabulated
expressions for the rate of heating $\Sdot$, the rate of change in
electron abundance $d\Ye/dt$, and the rate of change in the mean
atomic mass $d\Abar/dt$ in terms of the reaction rate for the
$\carbon+\carbon$ reaction.  These expressions are useful input for
large-scale hydrodynamical simulations of the simmering phase which
do not resolve such microphysics.

We first give, in \S~\ref{s.maxYe}, a simple estimate for the reduction in \Ye\ during the pre-explosion simmering and describe the role of \neon[22] and other trace nuclides.  In \S~\ref{s.numerical-results} we describe our numerical calculations, explain the reactive flows that occur (\S~\ref{s.flows}), and give simple approximations to the heating rate and carbon consumption (\S~\ref{s.effective-Q}). We detail, in \S~\ref{s.integration}, some of the limitations of our approach.  We evaluate the energy required to raise the white dwarf central temperature, and hence the amount of \carbon\ that must be consumed, and compare it against the one-zone calculation. We also estimate the central temperature at which convective mixing becomes faster than electron captures. This convective mixing advects
electron capture products to lower densities where they can
$\beta^{-}$-decay: the convective Urca process. Because each electron
capture-decay cycle emits a neutrino--anti-neutrino pair, there is
energy lost from the white dwarf, and our calculation underestimates
the amount of \carbon\ consumed prior to the flame runaway. The
convective Urca process \citep{Paczynski1972Carbon-Ignition} reduces
the rate of heating by nuclear reactions (thereby increasing the
amount of \carbon\ that must be consumed to raise the temperature), but
cannot result in a net decrease in entropy and temperature for
constant or increasing density
\citep{Stein1999The-Role-of-Kin,Stein2006The-Convective-}.  The Urca
reactions also tend to reduce the effects of buoyancy, and in
degenerate matter have a direct influence on the convective velocity
\citep{Lesaffre2005A-two-stream-fo}.  The paper concludes  (\S~\ref{s.discussion})  with a discussion of the implications for systematic variations in the mass of \nickel[56] produced in the explosion. 

\section{The reduction in electron abundance during the explosion}\label{s.maxYe}

The demise of an accreting white dwarf begins when the central temperature and density are such that the heating from the $\carbon+\carbon$ reaction becomes greater than the cooling from thermal neutrino emission.  For a density $\rho = 2.0\ee{9}\nsp\grampercc$ this requires a temperature $T \approx 3.0\ee{8}\nsp\K$ \citep[see][for a recent calculation]{Gasques2005Nuclear-fusion-}.  Initially the heating timescale is long, $\tH \equiv T(d T/d t)^{-1}\sim 10^{3}\nsp\yr$; as the temperature rises and the reaction rate increases, $\tH$ decreases. \citet{Woosley2004Carbon-Ignition} estimate that when $T > 7.6\ee{8}\nsp\K$, fluctuations in the temperature are sufficient to ensure that a local patch can run away and the flame ignites.

The basic reactions during \carbon\ burning were first worked out in
the context of core carbon burning in evolved stars
\citep{Reeves1959Nuclear-Reactio,Cameron1959Carbon-Thermonu}. During
simmering, \carbon\ is primarily consumed via
$\carbon(\carbon,\alpha)\neon$ and
$\carbon(\carbon,\pt)\sodium$. These reactions occur with a branching
ratio $0.56/0.44$ for $T < 10^{9}\nsp\K$ \citep{caughlan88:_therm}.
At temperatures below $\approx 7\ee{8}\nsp\K$, neutronization---that
is, a reduction in \Ye---occurs via the reaction chain
$\carbon(\pt,\gamma)\nitrogen[13](e^{-},\nu_{e})\carbon[13]$. This
electron capture implies that there is a maximum \Ye\ at the time of
the explosion, as first pointed out by
\citet{Piro2007Neutronization-}.  One can readily estimate the change
in electron abundance, $\Delta\Ye$. For simplicity, take the branching
ratio for $\carbon+\carbon$ to be 1:1 for producing $\pt+\sodium[23]$
and $\helium+\neon$.  Thus there is one \pt\ produced for every four
\carbon\ consumed. Two additional \carbon\ are consumed via
$\carbon(\nt,\gamma)\carbon[13](\helium,\nt)\oxygen$, which also
destroys one \helium\ nucleus, and via
$\carbon(\pt,\gamma)\nitrogen[13](e^{-}, \nu_{e})\carbon[13]$, which
also destroys one \pt.  Thus for every 6 \carbon\ consumed there is
one electron capture, so that $d\Ye/d Y_{12}\approx 1/6$, where $Y_{12}$ is
the molar abundance of \carbon.  As an
estimate for the heating from this reaction sequence, summing over the
$q$-values for the strong reactions gives a net heat release of
$(16\nsp\MeV)/6 = 2.7\nsp\MeV$ per \carbon\ nucleus consumed. Note
that at densities above $1.7\ee{9}\nsp\grampercc$, the reaction
$\sodium[23](e^{-},\nu_{e})\neon[23]$ contributes to the rate of
decrease in \Ye, so that $d\Ye/d Y_{12}\approx 1/3$ at those densities.  The total effective rate of $d\Ye/dY_{12}$ depends on the rate of convective mixing and the size of the convective core (see \S~\ref{s.integration}) but is always at least as large as the contribution from $\nitrogen[13](e^{-}, \nu_{e})\carbon[13]$.  In the following sections, we investigate these reactions
in detail.

\subsection{The role of neon-22 and other trace nuclides}\label{s.ne22-role}

Reactions on \neon[22], \sodium[23], and other trace nuclides also occur
during shell-burning in asymptotic giant branch (AGB) stars, and we briefly summarize their role in that context before describing the very different environment in a simmering white dwarf core.  In AGB
stars more massive than about $4\nsp\Msun$, the hydrogen
burning shell, at a temperature of 60-100\nsp\Mega\K, extends into
the convective envelope.  The envelope composition is then directly
affected by the various hydrogen-burning cycles: CNO, NeNa, and MgAl
\citep{Lattanzio:1997qk, Herwig2005Evolution-of-As}.  The hydrogen-burning
shell is also disturbed by thermal pulses due to ignition of the
helium layer. At each pulse, dredge-up of material may occur, in which
helium-burnt material is mixed into the stellar envelope, polluting it
with \helium, \carbon, \neon[22], and heavy $s$-process elements
\citep{Izzard2007Reaction-rate-u}.  Thus, the fate of \neon[22] is either to
contribute to hydrogen burning via $\neon[22](\pt,\gamma)\sodium[23]$ or
to become a neutron source for the $s$-process via the
$\neon[22](\alpha,\nt)\magnesium[25]$ reaction. The $\neon[22](\alpha,\nt)\magnesium[25]$ reaction requires the high
temperatures ($T > 2.5\ee{8}\nsp\K$) that can be found at the
bottom of the pulse-driven convective zone during the helium shell
flashes. The neutrons are released with high density [$\log (N_n/\cm^{-3}) \sim
10$] in a short burst \citep{Gallino:1998ax, Busso1999Nucleosynthesis}. These
peak neutron densities are realized for only about a year, followed by
a neutron density tail that lasts a few years, depending on the
stellar model assumptions.  These neutrons are the genesis of the
classic high-temperature $s$-process in AGB stars.

Should \sodium[23] be present, there are two usual possibilities for
subsequent nuclear processing in AGB stars via either the
$\sodium[23](\pt,\alpha)\neon$ reaction or the
$\sodium[23](\pt,\gamma)\magnesium$ reaction. The former reaction gives rise to
the classical NeNa cycle \citep{Marion:1957ox,Rolfs:1988jr,Rowland:2004dp}, whereas
the competing $(p,\gamma)$ reaction transforms \sodium[23] to heavier isotopes and
bypasses the NeNa cycle.  How much material is processed through the
$(\pt,\alpha)$ reaction on \sodium[23] as opposed to the competing $(\pt,\gamma)$ reaction is of interest for AGB star (and classical novae)
nucleosynthesis.  New measurements of the $\sodium[23]+\pt$ cross section \citep{Rowland:2004dp} suggest that for $T=(20\textrm{--}40)\nsp\Mega\K$ $\sodium[23](\pt,\gamma)\magnesium[24]$ competes with the $(\pt,\alpha)$ branch, disrupts the NeNa cycle, and produces a flow into the MgAl hydrogen burning cycle. 

\begin{figure}[htbp]
\includegraphics[width=86mm]{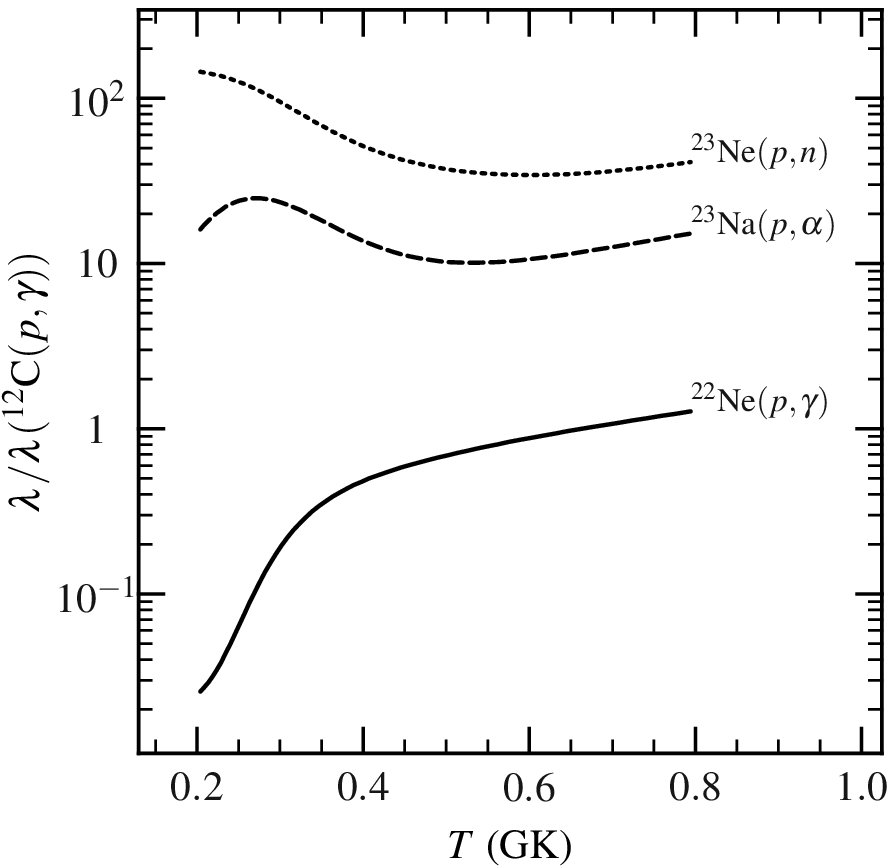}
\caption{Ratio of thermally averaged cross-sections, $\lambda \equiv \NA\langle\sigma v\rangle$, to that for $\carbon(\pt,\gamma)$, for three reactions: $\neon[22](\pt,\gamma)$ (\emph{solid line}); $\sodium[23](\pt,\alpha)$ (\emph{dashed line}); and $\neon[23](\pt,\nt)$ (\emph{dotted line}). Screening is included in $\lambda$; we evaluate the ratio at $\rho = 3\ee{9}\nsp\grampercc$ for $\neon[23](\pt,\nt)$ and at $10^{9}\nsp\grampercc$ for the other two.  In the temperature range where the heating timescale is slow enough for weak interactions to reduce \Ye\ (cf.\ Fig.~\ref{f.dYe}), the thermally averaged cross-sections for $\neon[22](\pt,\gamma)$ and $\carbon(\pt,\gamma)$ are of similar magnitude. For \neon[22] to compete with \carbon\ for \pt-captures at $T \gtrsim 3\ee{8}\nsp\K$ requires a \neon[22] mass fraction $X_{22} \gtrsim (22/12) X_{12}$. At densities less than the electron capture threshold on \sodium[23] a small flow of $\sodium[23](\pt,\alpha)\neon[20]$ can occur (cf.\ Fig.~\ref{f.flow0rho9}).  At higher densities \sodium[23] electron captures to form \neon[23]; the large  cross-section for $\neon[23](\pt,\nt)\sodium[23]$ allows it to compete with captures onto \carbon\ if $Y_{23}/Y_{12} \gtrsim 1\%$. }
\label{f.screened-rates-compare}
\end{figure}

Caution about intuition developed for reaction sequences on \neon[22] and
\sodium[23] in AGB star environments seems prudent when applied to the
dense, carbon-rich environments of white dwarfs near the Chandrasekhar
mass limit.
During the slow \carbon\ simmering preceding the explosion, the large $Y_{12}$ ensures that \pt\ liberated by the \carbon(\carbon,\pt)\sodium[23] branch will capture preferentially onto \carbon, rather than \neon[22] or \sodium[23].  Figure~\ref{f.screened-rates-compare} shows the ratio of thermally averaged cross-sections, $\lambda \equiv \NA\langle\sigma v\rangle$, to that for $\carbon(\pt,\gamma)\nitrogen[13]$ for three reactions: $\neon[22](\pt,\gamma)\sodium$ (\emph{solid line}), $\sodium(\pt,\alpha)\neon[20]$ (\emph{dashed line}), and $\neon[23](\pt,\nt)\sodium$ (\emph{dotted line}).  In addition to the Coulomb penetration, there are numerous resonances that determine how the cross-sections change with temperature.  When the ratio of the thermally averaged cross-sections is of order unity, as it is for $\neon[22](\pt,\gamma)\sodium[23]$, then the proton capture is determined by the relative abundances of \neon[22] and \carbon.

Note that for the latter two reactions, we plot the largest
proton-consuming branch rather than the $(\pt,\gamma)$ branch.  We include
screening in all reactions \citep{Yakovlev2006Fusion-reaction} with
the plasma taken to consist of \carbon\ and \oxygen\ with mass
fraction 0.3 and 0.7, respectively. For $T > 10^{8}\nsp\K$,
$\lambda[\neon[22](\pt,\gamma)\sodium]$ is well-constrained
experimentally \citep{Iliadis2001Proton-induced-}. For a white dwarf
with a central density $\rho = 2\ee{9}\nsp\grampercc$ the ignition
temperature (where nuclear heating dominates over cooling via thermal
neutrino emission) is $\approx 3\ee{8}\nsp\K$
\citep{Gasques2005Nuclear-fusion-}; the burning timescale becomes less
than the timescale for $\sodium[23](e^{-},\nu_{e})\neon[23]$ once the temperature
increases beyond $T \gtrsim 6\ee{8}\nsp\K$.  Over this range, the
thermally averaged cross-sections for
$\neon[22](\pt,\gamma)\sodium[23]$ and
$\carbon(\pt,\gamma)\nitrogen[13]$ are comparable, but the abundance
of \carbon\ is far greater; having \pt\ capture preferentially onto
\neon[22] would therefore require it to have a mass fraction roughly
twice that of \carbon.

At densities greater then $1.7\ee{9}\nsp\grampercc$, the reaction
$\sodium[23](e^{-},\nu_{e})\neon[23]$ produces roughly one \neon[23]
nucleus for every six \carbon\ nuclei consumed.  The screened thermally
averaged cross-section for $\neon[23](\pt,\nt)\sodium[23]$ is $\gtrsim
30$ times that of $\carbon(\pt,\gamma)\nitrogen[13]$
(Fig.~\ref{f.screened-rates-compare}), so that \neon[23] could become
a competitive sink for protons. For our self-heating burn (see
\S~\ref{s.flows}) starting at $\rho = 3\ee{9}\nsp\grampercc$, the
abundance of \neon[23] reaches $Y_{23} = 3\ee{-4}$ ($Y_{23}\approx
0.015Y_{12}$) by the point the heating timescale $\tau_{\mathrm{H}}$
becomes shorter than the timescale for electron captures onto \sodium[23]. Although our one-zone approximation overestimates the amount of \carbon\ that must be consumed to raise the central temperature (see \S~\ref{s.integration}), should enough of the convective core lie above the threshold for electron capture onto \sodium, the abundance of \neon[23] can become large enough to choke off
the $\carbon(\pt,\gamma)\nitrogen[13]$ reaction, as noted by \citet{Piro2007Neutronization-}.

\subsection{The reaction $\nitrogen[13](e^{-},\nu_{e})\carbon[13]$}\label{s.estimates}

As described above, the large \carbon\ abundance ensures that protons
produced by $\carbon(\carbon,\pt)\sodium[23]$ lead to the formation of
$\beta^{+}$-unstable \nitrogen[13] via
$\carbon(\pt,\gamma)\nitrogen[13]$ unless an appreciable abundance of
\sodium[23] or \neon[23] can build up.  At these densities, the rate
for electron capture is substantially greater than the rate of
$\beta^{+}$-decay for \nitrogen[13]. The electron Fermi energy is
$\approx 5.1\nsp\MeV(\rho\Ye/10^{9}\nsp\grampercc)^{1/3}$, and the $q$-value
for the $\beta^{+}$ decay of \nitrogen[13] is $2.2\nsp\MeV$. As a
result, there are several excited states of \carbon[13] into which the
electron can capture.  Of these, the transition to the excited state
$E_{\mathrm{exc}}=3.68\nsp\MeV$, with spin and parity
$J^{\pi}=3/2^{-}$, is an allowed Gamow-Teller transition.  We computed
the electron capture rate using the experimental $\log ft$ for the
ground-state transition
\citep{Ajzenberg-Selove1991Energy-levels-o}. Gamow-Teller strengths to
excited states were calculated with the shell-model code \code{OXBASH}
\citep{Brown2004OXBASH-for-Wind} employing the Cohen-Kurath II
potential \citep{Cohen1967Spectroscopic-f} in the $p$-shell model
space. A quenching factor of 0.67 \citep{Chou1993Gamow-Teller-be} was
applied to this strength, and the resulting $ft$ values were used with
the analytical phase space approximations of
\citet{Becerril-Reyes2006Electron-Captur} to obtain the capture rate.
These shell-model calculations agree well with recent $(\helium[3],t)$
scattering data \citep{Zegers2007Gamow-Teller-St}. At $\rho\Ye =
10^{9}\nsp\grampercc$, captures into the excited state at
$E_{\mathrm{exc}}=3.68\nsp\MeV$ account for $\gtrsim 0.3$ of the total
rate ($R_{\mathrm{ec}} = 12\nsp\second^{-1}$); this fraction increases
with density. Because $R_{\mathrm{ec}} > (G\bar{\rho})^{-1/2}$, this
capture does not freeze out during the simmering, unlike capture onto
\sodium[23].  Although the capture into the excited level does not
increase the capture rate substantially beyond that for the
ground-state-to-ground-state transition, it does increase the heat
deposited into the white dwarf from this reaction. We find the heat
deposited from this reaction, at $\rho\Ye = 10^{9}\nsp\grampercc$, to
be $1.3\nsp\MeV$.

\subsection{Production and subsequent captures of neutrons}\label{s.neutron-captures}

Finally, we consider the contribution from heavier nuclei, such as \iron, inherited from the main-sequence star. In AGB stars the reaction $\carbon[13](\alpha,\nt)\oxygen$ (during a He shell flash, $\neon[22](\alpha,\nt)\magnesium[25]$ also contributes) acts as a neutron source for the \emph{s}-process. In contrast, the large \carbon\ abundance of the white dwarf core prevents a strong \emph{s}-process flow during the pre-explosion simmering. The cross-section for  $\carbon[12](\nt,\gamma)\carbon[13]$ is 63.5 times smaller than the cross-section for $\iron[56](\nt,\gamma)\iron[57]$ \citep{Bao1987Neutron-capture} at an energy of 30\nsp\MeV, which is not sufficient to overcome the vastly larger abundance of \carbon\ nuclei (for a progenitor with solar metallicity, the \carbon:\iron\ ratio [for a \carbon\ mass fraction of 0.3] is 1400:1).

\section{Reaction network calculations}\label{s.numerical-results}

In this section we investigate the reactions that occur during simmering in more detail using a ``self-heating'' reaction network. Under isobaric conditions the temperature $T$ evolves with time according to
\begin{equation}\label{eq:T}
\frac{dT}{dt} =  \frac{\Sdot}{C_{P}}.
\end{equation}
Here $C_{P}$ is the specific heat at constant pressure, and the heating rate \Sdot\ is given by
\begin{equation}\label{eq:Sdot}
\Sdot = - \NA \left[ c^2 \sum_{i} \left(M_{i}\frac{dY_{i}}{dt}\right)  + \mu_{e} \frac{d\Ye}{dt}  \right]- \Sdot_{\nu} ,
\end{equation}
where $M_{i}$ and $Y_{i}$ are, respectively, the atomic mass and
molar abundance of species $i$, $\mu_{e}$ is the electron chemical potential, $\Sdot_{\nu}$ is the neutrino loss rate, per unit mass, from the weak reactions \citep{Fuller1982Stellar-weak-in,langanke.martinez-pinedo:weak}, and
$\NA = 6.022\ee{23} \nsp\gram^{-1} = (1\nsp\mathrm{amu})^{-1}$. We neglect thermal neutrino emission processes; this is an excellent approximation over most of the integration. Our reaction network incorporated 430 nuclides up
to \germanium[76] and is the same one used to explore the effect of
\neon[22] on the laminar flame speed \citep[see][and references
therein, for a description of the
microphysics]{Chamulak2007The-Laminar-Fla}. At conditions of
$\rho=(1\textrm{--}3)\ee{9}\nsp\grampercc$ and $T = 5\ee{8}\nsp\K$,
the specific heat $C_{P}$ is dominated by the ions, which are in a
liquid state (plasma parameter $\Gamma \equiv \langle Z^{5/3}\rangle
\left( e^{2}/\kB T \right) \left(\rho\Ye\NA\right)^{1/3} \approx 10$),
and have $C_{P} \approx 2.9\kB\NA/\langle A\rangle$,
where $\langle A\rangle$ is the average atomic mass.

\subsection{The reactive flows}\label{s.flows}

In this section we refine our estimate of $d\Ye/d Y_{12}$ made in
\S~\ref{s.estimates}.  We integrate equations
(\ref{eq:T})--(\ref{eq:Sdot}) starting from the temperature at which
heating from the $\carbon+\carbon$ reaction equals the heat loss
from thermal neutrino losses (this determines the onset of thermal
instability). For simplicity, we split the solution of the thermal and
network equations. That is, for each time-step $dt$ we solve the
thermal equations to obtain $T$ and $\rho$, integrate the reaction
network at that $T$ and $\rho$ to compute $Y_{i}$ and \Sdot, and use
\Sdot\ to advance the solution of equation (\ref{eq:T}).

In the initial phases of the simmering, the convective timescale is
slow, and our one-zone calculations give an adequate description of
the heating (when corrected for the gradient in temperature).  As the
temperature of the white dwarf increases, the heating timescale \tH\
decreases; moreover, the convective mixing becomes more rapid, and one
must treat the hydrodynamical flows in order to calculate the
nucleosynthesis properly (see \S~\ref{s.integration}).  In this
section, we will restrict our integration to where $T <
0.6\nsp\Giga\K$, for which the heating timescale is $\gtrsim
10^{4}\nsp\second$, so that electron captures onto \sodium[23] are not
frozen out.

\begin{figure}[htbp]
\includegraphics[width=86mm]{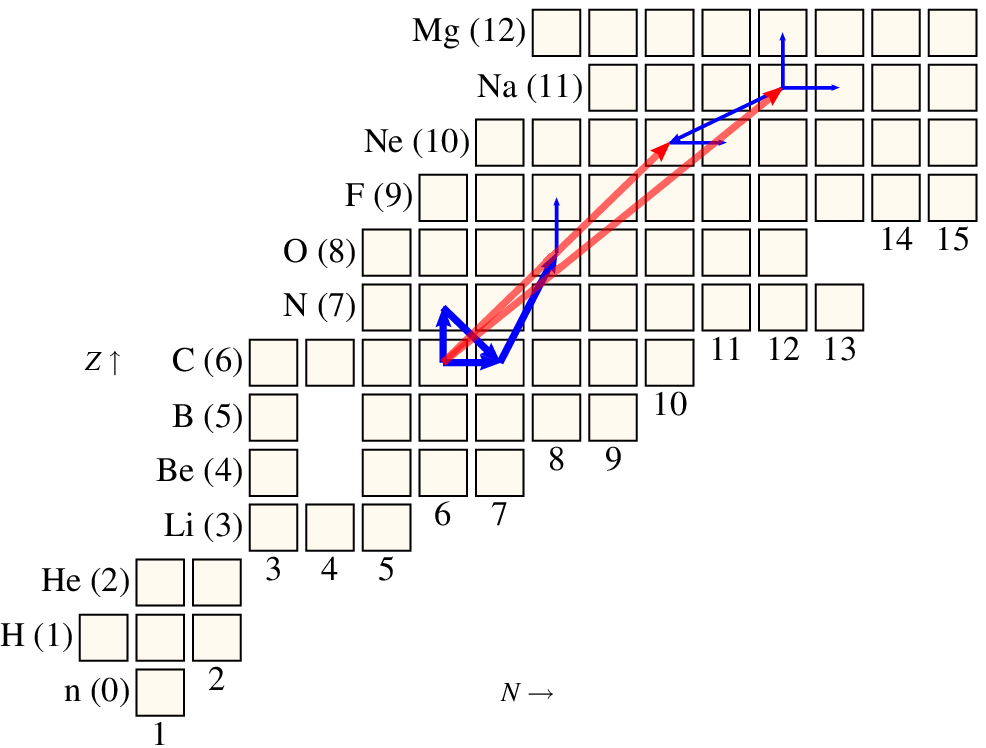}
\caption{Flows during a constant-pressure self-heating burn at $\rho = 10^{9}\nsp\grampercc$ with an initial composition $\{X_{12},X_{16},X_{22}\} = \{0.3,0.7,0.0\}$.  We integrate over the period of the burn with $T < 6\ee{8}\nsp\K$. Each row contains the isotopes of a particular element $Z$, with the columns containing different neutron numbers $N$. The width of the arrow is proportional to the magnitude of the flow, with only those flows having magnitude $> 0.01\cdot\max(F)$ being shown. The primary $\carbon+\carbon$ reactive flows are indicated with a lighter shading and are in red in the online version. }
\label{f.flow0rho9}
\end{figure}

\begin{figure}[htbp]
\includegraphics[width=86mm]{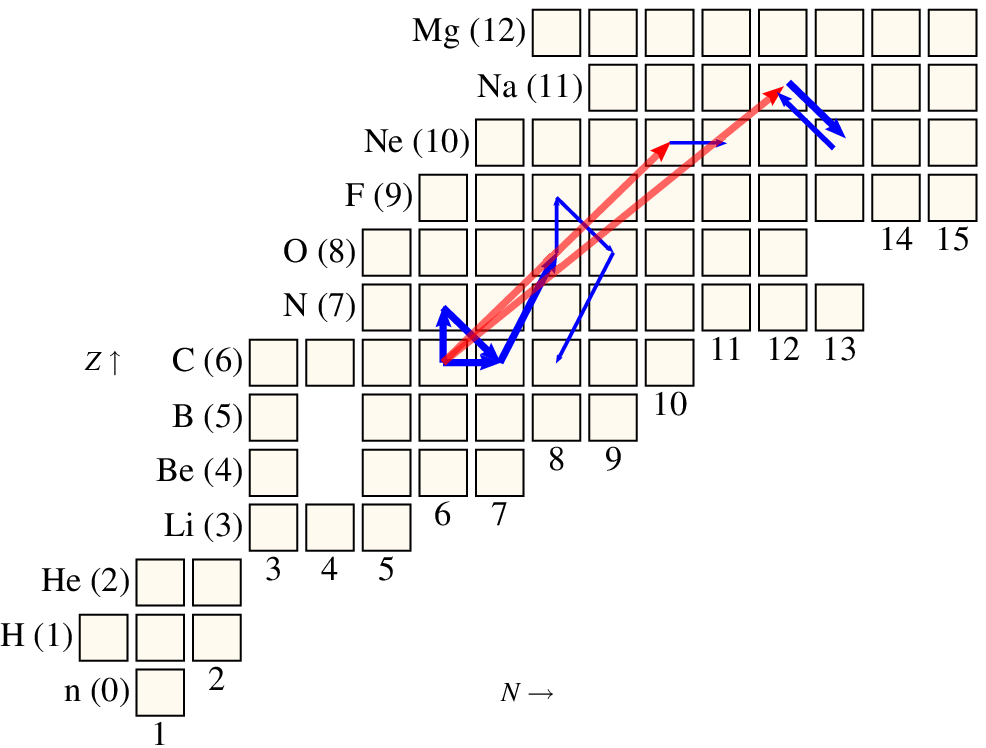}
\caption{Same as in Fig.~\ref{f.flow0rho9}, but at $\rho = 3\ee{9}\grampercc$.  The reaction $\sodium(e^{-},\nu_{e})\neon[23]$ is now the dominant destroyer of \sodium\ instead of $(\pt,\gamma)$, $(\pt,\alpha)$, and $(\nt,\gamma)$ reactions.}
\label{f.flow0}
\end{figure}

To explore the reaction channels that link \carbon\ consumption with the reduction in \Ye, 
we define the reactive flow between nuclides $i$ and $j$ as
\begin{equation}
  F\left(i\to j\right) \equiv \int_{t}\,\left|\frac{d Y_{i}}{d t}\right|_{i\to j}\,d t ,
\end{equation}
where the integral is for the reactions linking nuclide $i$ with $j$.  We differentiate between $F(i\to j)$ and $F(j\to i)$, i.e., we treat inverse rates separately. Figures~\ref{f.flow0rho9} and \ref{f.flow0} show the reactive flows for $\rho = 10^{9}\nsp\grampercc$ and $3\ee{9}\nsp\grampercc$, respectively. In both cases the composition is $\{X_{12}, X_{16}, X_{22}\} = \{0.3,0.7,0.0\}$.  Each row of the chart is a different element ($Z$), with the columns corresponding to neutron number. For viewing ease, we only plot those flows having $F > 0.01\cdot\max(F)$, and we indicate the strength of the flow by the line thickness.  We highlight the flows of the initial \carbon\ fusion reactions with a lighter shading (these flows are also in red in the online version). For illustrative purposes, we only show the flows for when the temperature is not sufficiently hot for photodissociation of \nitrogen[13]. Finally, by convention and to avoid cluttering the plot, we do not show flows into or out of \pt, \nt, and \helium.

To facilitate comparisons between different runs, we define a normalized flow as
\begin{equation}\label{e.normalizedF}
  \Fn(i\to j) \equiv \frac{F(i\to j)}{F(\carbon\to \sodium) + F(\carbon\to \neon)}.
\end{equation}
At a density of $10^{9}\nsp\grampercc$ we find $\Fn(\carbon\to\sodium[23]) = 0.43$ and $\Fn(\carbon\to\neon[20]) = 0.57$, which reflects the branching ratio \citep{caughlan88:_therm}. In agreement with the arguments in \S~\ref{s.maxYe}, most of the protons liberated by $\carbon(\carbon,\pt)\sodium$ capture onto \carbon, with $\Fn(\carbon\to\nitrogen[13]) \approx  \Fn(\nitrogen[13]\to\carbon[13]) = 0.20$. Note that if all \pt\ were to capture onto \carbon, we would have $\Fn(\carbon\to\nitrogen[13])/\Fn(\carbon\to\sodium) = 0.5$, because the reaction $\carbon(\carbon,\pt)\sodium$ consumes two \carbon.    
The $\alpha$-particle released by the $\carbon(\carbon,\alpha)\neon$ reaction captures onto \carbon[13] to form \oxygen\ and a neutron, which in turn destroys another \carbon\ via $\carbon(\nt,\gamma)\carbon[13]$.  The flow $\Fn(\carbon\to\carbon[13]) = 0.26$ and nearly matches the number of $\alpha$-particles produced by $\carbon(\carbon,\alpha)\neon$; \nt-captures onto \sodium\ and \neon\ account for the difference.  At $\rho=10^{9}\nsp\grampercc$ the only electron captures are onto \nitrogen[13], so $\Delta \Ye = -\Fn(\nitrogen[13]\to\carbon[13])$.  Dividing by $\Delta Y_{12}=-[\Fn(\carbon\to\sodium)+\Fn(\carbon\to\neon)+\Fn(\carbon\to\nitrogen[13])+\Fn(\carbon\to\carbon[13])]$ gives $d \Ye / d Y_{12} = 0.14$ (Table \ref{t:Yechange}). This is slightly less than the estimate of $1/6$ (\S~\ref{s.estimates}) because of the lower branching ratio of $\carbon\to\sodium$.

At $\rho = 3\ee{9}\nsp\grampercc$, \sodium[23] is consumed by the reaction $\sodium[23](e^{-},\nu_{e})\neon[23]$ rather than by \pt- or \nt-capture (cf.\ Figs.~\ref{f.flow0rho9} and \ref{f.flow0}).  The flow is much larger than that from \nitrogen[13], because the reaction $\neon[23](\pt,\nt)\sodium[23]$ competes for \pt\ and produces more \sodium[23].  Indeed, $\Fn(\carbon\to\carbon[13]) = 2.5 \Fn(\carbon\to\nitrogen[13])$ because of extra \nt\ produced by the reaction $\neon[23](\pt,\nt)\sodium[23]$.  There is an additional contribution from electron captures onto \fluorine[17] produced via $\oxygen(\pt,\gamma)\fluorine[17]$, but this flow is only about 4\% of the $\sodium[23](e^{-},\nu_{e})\neon[23]$ flow. At both densities, \neon[22] plays a small role in reducing \Ye, which we verified by comparing the flows for a burn with $X_{22} = 0.06$ with those for a flow with $X_{22} = 0.0$ (Fig.~\ref{f.flow0}).  We find that, for a burn starting at $\rho = 3.0\ee{9}\nsp\grampercc$, $\Fn(\neon[22]\to\sodium[23])/\Fn(\carbon\to\nitrogen[13]) = 0.08$ and $\Fn(\neon[22]\to\neon[23])/\Fn(\carbon\to\carbon[13]) = 0.05$.  Although  $\lambda[\neon[22](\pt,\gamma)\sodium[23]]\lesssim\lambda[\carbon[12](\pt,\gamma)\nitrogen[13]]$ at $T\lesssim 6\ee{8}\nsp\K$ (Fig.~\ref{f.screened-rates-compare}), the abundances are in ratio $Y_{22}/Y_{12} = 0.11$. Below the electron capture threshold for \sodium[23], the reaction $\neon[22](p,\gamma)\sodium[23]$ will cause a slight reduction in $d\Ye / d Y_{12}$ equal to the ratio of $\Fn(\neon[22]\to\sodium[23])$ to $\Fn(\carbon\to\nitrogen[13])$.

Finally, we confirmed the lack of an s-process flow (\S~\ref{s.neutron-captures}) by performing a run with nuclides from \neon\ to \iron\ present at up to 3 times their solar abundances \citep{anders.grevesse:abundances}.  The threshold for electron capture onto \iron\ is $1.5\ee{9}\nsp\grampercc$, and so at higher densities carbon ignition occurs in a more neutron-rich environment.  We therefore start the calculation by artificially suppressing the strong interactions and allowing the mixture to come into $\beta$-equilibrium.  We then turn on the strong reactions and let the runaway commence. In all cases the heavier nuclides did not have a substantial impact on the reactive flows.  The value for $d \Ye / d Y_{12}$ is somewhat larger than 0.3 at densities $\rho \ge 1.2\ee{9}\nsp\grampercc$, the threshold density for $\magnesium[25](e^{-},\nu_{e})\sodium[25](p,n)$, because of the reactions $\magnesium[24](\pt,\gamma)\aluminum[25](e^{-}, \nu_{e})\magnesium[25](e^{-},\nu_{e})\sodium[25]$. Of these two captures, \aluminum[25] is $\beta^{+}$-unstable, and hence the electron capture is fast enough to proceed throughout simmering; the capture onto \magnesium[25] has a timescale, at $\rho = 2\ee{9}\nsp\grampercc$, of $\approx 900\nsp\second$ and will therefore freezeout during simmering, just as captures onto \sodium[23] freezeout.

Our runs span a range of initial densities, from $10^{9}\nsp\grampercc$ (for which the electron Fermi energy is too low to induce electron captures onto \sodium[23]) to $6\ee{9}\nsp\grampercc$, which represents an extreme case for accretion onto a cold, initially massive white dwarf \citep{Lesaffre2006The-C-flash-and}. In all cases we took the initial \carbon\ mass fraction to be $X_{12} = 0.3$. As noted above, the mass fraction of \neon[22] would have to exceed that of \carbon\ to change the nucleosynthesis during simmering appreciably. At higher densities, $\pt$-captures onto \magnesium\ can also play a role (\S~\ref{s.flows}), but our results will not change appreciably so long as $X_{12}$ is not substantially less than 0.3.

The calculation, being a single reaction network integration, does not
incorporate the effects of mixing in the white dwarf core. Our focus
here is to elucidate the reactions that set \Ye. These calculations do
not determine the total amount of carbon consumed (although see
\S~\ref{s.integration} for an estimate) or the total mass of processed
material that lies at a density greater than the electron capture
threshold. We list our one-zone results in terms of the change in
electron abundance per carbon consumed, $d\Ye/d
Y_{12}$. Table~\ref{t:Yechange} summarizes our numerical findings of
$d\Ye / d Y_{12}$ for densities $10^{9}$, $3\ee{9}$, and
$6\ee{9}\nsp\grampercc$, for \neon[22] mass fractions $X_{22}=0$ and
0.06, and finally a run (denoted as $3Z_{\sun}$ in
Table~\ref{t:Yechange}) with elements heavier than \neon\ present at
3 times solar abundance. We use this value of $3Z_{\sun}$ as
representing a rough upper limit based on the $\approx
0.5\nsp\unitstyle{dex}$ scatter in $[\mathrm{Fe/H}]$ present in local
field stars \citep{feltzing.holmberg.ea:solar}.

\begin{deluxetable}{cccc}
\tablecolumns{8}
\tablecaption{Change in electron abundance per carbon consumed during the pre-explosion convective burning
\label{t:Yechange}}
\tablewidth{0bp}
\tablehead{\colhead{composition\tablenotemark{a}} & \colhead{density} & \colhead{$d \Ye / d Y_{12}$ } & \colhead{$\sum_{i} dY_i / d Y_{12}$} \\
& \colhead{$10^9$\nsp\grampercc} & }
\startdata
  $X_{22} = 0.00$ & 1.0 & 0.136 & 0.340\\
  $\cdots$ & 3.0 & 0.297 & 0.340 \\
  $\cdots$ & 6.0 & 0.302 & 0.342\\ 
  $X_{22} = 0.06$ & 1.0 & 0.125 & 0.347\\
  $\cdots$ & 3.0 & 0.301 & 0.344\\
  $\cdots$ & 6.0 & 0.305 & 0.346\\
  $3 Z_\odot$ & 1.0 &  0.130 & 0.361\\
  $\cdots$ & 3.0 & 0.349 & 0.370\\
  $\cdots$ & 6.0 & 0.355 & 0.380\\
\enddata
\tablenotetext{a}{In all cases the initial mass fraction of \carbon\ is 0.3.}
\end{deluxetable}

\subsection{The effective $q$-value of the $\carbon+\carbon$ reaction}\label{s.effective-Q}

The scope of this work is to elucidate the nuclear reactions that
occur during the pre-explosion simmering, and including their effects
in large-scale hydrodynamics simulations of the entire white dwarf is
advisable.  As an aid to such simulations, we present fits for the
carbon depletion rate and effective heat deposition, which improve
on previous expressions \citep{Woosley2004Carbon-Ignition}. Since the
reaction chain is controlled by the reaction $\carbon +\carbon$, we
write the rate of carbon consumption, $d Y_{12}/d t$ as being
proportional to the thermally averaged cross-section, $\lambda =
\NA\langle \sigma v\rangle$
\begin{equation}\label{e.multiplicity-def}
\frac{d Y_{12}}{d t} = -M_{12} \left(\frac{1}{2}Y_{12}^{2} \,\rho \lambda\right).
\end{equation}
This definition is such that the quantity in parenthesis is the reaction rate per pair of \carbon\ nuclei and $M_{12}=2$ if the only \carbon-destroying reaction present is $\carbon+\carbon$.  One can determine $M_{12}$ from summing the normalized reaction flows (eq.~[\ref{e.normalizedF}])  out of \carbon.

In a similar fashion, we can define the effective heat evolved, \qeff, per reaction $\carbon+\carbon$ by the equation
\begin{equation}\label{e.qeff-def}
\Sdot = \qeff\NA \left(\frac{1}{2} Y_{12}^{2}\,\rho\lambda\right).
\end{equation}
With this definition, one has $\Sdot = (\qeff\NA/M_{12})\times d Y_{12}/d t$; this may be compared with equation~(\ref{eq:Sdot}) to relate \qeff\ to the binding energy of the nuclei. To compute these quantities, we integrated the reaction network over a grid of $\rho$ and $T$, both of which were held fixed for each run.  We found in all cases that the instantaneous energy generation rate \Sdot\ would, after some initial transient fluctuations, settle onto a constant value until a significant ($> 10\%$) depletion in \carbon\ had occurred.  We used this plateau in \Sdot\ to obtain \qeff\ and $M_{12}$.  For the densities of interest, the values of both \qeff\ and $M_{12}$ obtained this way are nearly independent of temperature. We find $M_{12}=2.93$; this value is accurate to within 2\% over all our runs. The value of \qeff\ increases slightly with density, but is nearly constant over the temperature range of interest. We find that $\qeff = 8.91\nsp\MeV$, 9.11\nsp\MeV, and 9.43\nsp\MeV\ for $\rho = 10^{9}\nsp\grampercc$, $2\ee{9}\nsp\grampercc$, and $3\ee{9}\nsp\grampercc$, respectively.  At each of these densities, the quoted value of \qeff\ is accurate to within 3\% over the temperature range of $3\ee{8}\nsp\K$ to $7\ee{8} \nsp\K$.  

For use in large-scale hydrodynamic models, one first computes the
screened thermally averaged cross-section $\lambda$ \citep[for the
most recent rate, see][]{Gasques2005Nuclear-fusion-} at the
thermodynamic conditions of a given cell. Combining $\lambda$ and the
cell's carbon abundance $Y_{12}$ with our estimates of $M_{12}$ and
\qeff, one computes $dY_{12}/dt$ and \Sdot\ from
equations~(\ref{e.multiplicity-def}) and (\ref{e.qeff-def}).  In
effect, this procedure incorporates the results of a large reaction
network and careful treatment of the detailed nuclear physics into
simple expressions.
We caution that these fits were obtained in the regime $-\Delta Y_{12} < 0.003$, for which proton captures onto \sodium[23] are not competitive with proton captures onto \carbon. Note that the heat released per \carbon\ nucleus consumed is $\qeff/M_{12} \approx 3.1\nsp\MeV$, slightly higher than our simple estimate (\S~\ref{s.maxYe}), and also somewhat higher than those used by \citet{Piro2007Neutronization-}.  This is because of our inclusion of heating from the $\nitrogen[13](e^{-},\nu_{e})\carbon[13]$ and $\sodium[23](e^{-},\nu_{e})\neon[23]$ reactions. Our estimate of the heat evolved is less than that computed under the assumption that the products of \carbon\ burning are a 3:1 \neon:\magnesium\ mixture \citep[see, e.g.][]{Woosley2004Carbon-Ignition}, which releases 5.0\nsp\MeV\ per \carbon\ nucleus consumed\footnote{Note that in eq.~(1) of \citet{Woosley2004Carbon-Ignition}, the factor of \onehalf\ is subsumed into their quantity $\lambda_{12,12}$.}. Our net heating rate, per $\carbon+\carbon$ reaction, is about 10\% less than that used by \citet{Woosley2004Carbon-Ignition}, but we effectively consume $\approx 3$ \carbon\ nuclei per reaction.

The change in electron abundance is related to
$\lambda$ via
\begin{equation}\label{e.Ye-lambda}
\frac{dY_{e}}{dt} = -\left(M_{12} \frac{dY_{e}}{dY_{12}}\right)\times \left(\frac{1}{2} Y_{12}^{2}\,\rho\lambda\right),
\end{equation}
where $dY_{e}/dY_{12}$ is taken from Table~\ref{t:Yechange}.  
Finally, we may compute the change in the mean atomic mass, $\Abar$, as a function of carbon consumed.  On differentiating $\Abar = (\sum_{i} dY_{i}/d Y_{12})^{-1}$, and substituting equation~(\ref{e.multiplicity-def}, we have
\begin{equation}\label{e.dAbar}
\frac{d\Abar}{dt} = \Abar^{2}M_{12}\left(\frac{1}{2}Y_{12}^{2}\rho\lambda\right)\left(\sum_{i}\frac{dY_{i}}{dY_{12}}\right),
\end{equation}
where the quantity $\sum_{i} dY_{i}/dY_{12}$ is computed from the flows (eq.~[\ref{e.normalizedF}]) and is listed in Table~\ref{t:Yechange}. From the simple description of the reactions (\S~\ref{s.maxYe}; see also \citealt{Piro2007Neutronization-}) we have, for every six \carbon[12] destroyed, one each of \carbon[13], \oxygen, \neon, and \sodium[23] (or \neon[23]), so that $\sum_{i}dY_{i}/dY_{12} \approx\onethird$.  This results in a smaller change in \Abar\ than would result from burning $\carbon+\carbon$ to a 3:1 \neon:\sodium\ mixture (in that case, $\sum_{i}dY_{i}/dY_{12} = 0.42$).  We advocate using equations~(\ref{e.multiplicity-def})--(\ref{e.dAbar}) and the computed values (Table~\ref{t:Yechange}) of $d\Ye/dY_{12}$ and $\sum_{i} dY_{i}/dY_{12}$ in numerical simulations of simmering.

\subsection{Heating of the white dwarf core and the end of simmering}\label{s.integration}

\begin{figure}[htbp]
\includegraphics[width=86mm]{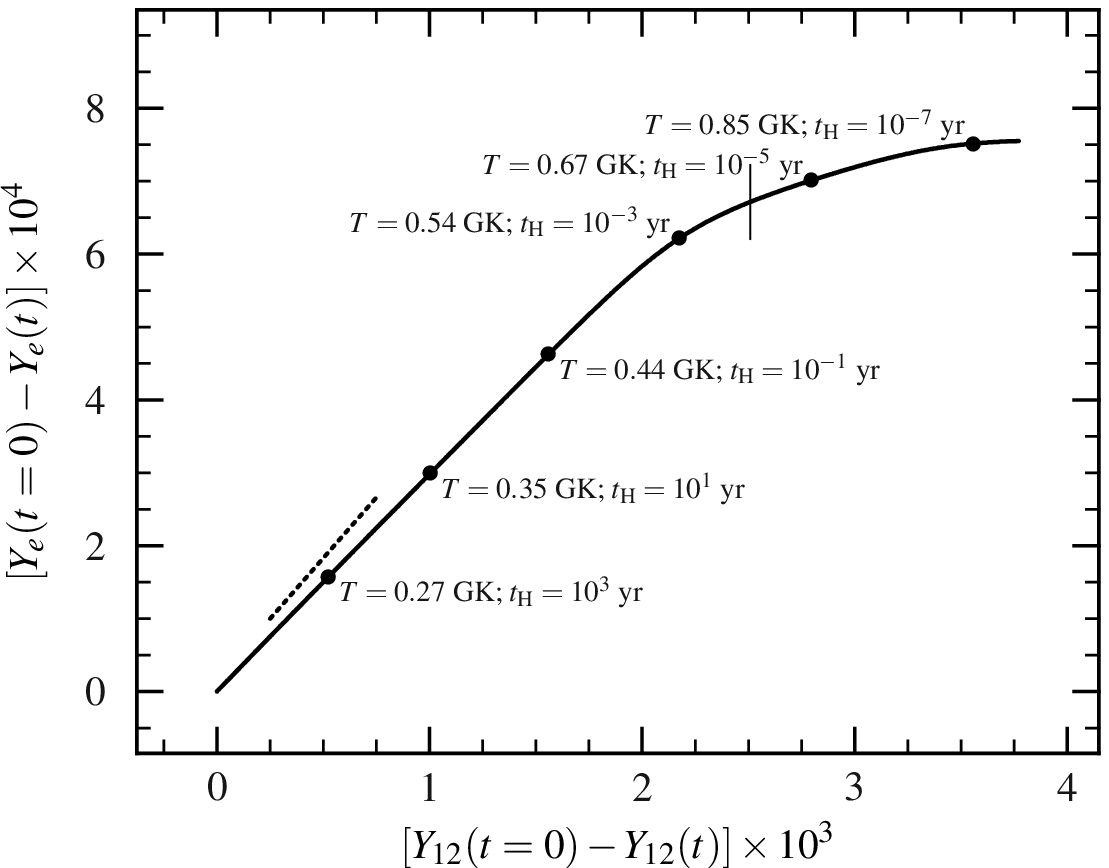}
\caption{Change in electron abundance, $\Ye(t=0)-\Ye(t)$, as a function of carbon consumed, $Y_{12}(t=0)-Y_{12}(t)$. The break in the slope, at $Y_{12}(0)-Y_{12}\approx 2\ee{-3}$, occurs when the heating timescale $\tH=C_{P}/\Sdot$ becomes less than the timescale for electron capture onto \sodium[23], which is $\approx 2700\nsp\second$ at $\rho = 3.0\ee{9}\nsp\grampercc$.  We indicate this point with the thin vertical line.  To guide the eye, the short dotted line indicates a slope of \onethird.}
\label{f.dYe}
\end{figure}

In previous sections, we evaluated the heating and neutronization of the white dwarf core in terms of the rate of \carbon\ consumption.  We now estimate the net amount of \carbon\ that is consumed in raising the white dwarf temperature and evaluate the temperature at which electron captures onto \sodium[23] freeze out. In the one-zone isobaric calculations, using equations~(\ref{e.multiplicity-def}) and (\ref{e.qeff-def}), we have $dT/dY_{12} = (\qeff\NA)/(M_{12} C_{P})$, so that $\Delta T \approx 0.15\nsp\Giga\K\left(\Delta X_{12}/0.01\right)$, where $\Delta X_{12}$ is the change in mass fraction of \carbon.  The change of \carbon\ abundance required to raise the temperature from $3\ee{8}\nsp\K$ to $8\ee{8}\nsp\K$ is then $|\Delta Y_{12}| \approx 2.8\ee{-3}$.  This is about 11\% of the available \carbon, for an initial \carbon\ mass fraction $X_{12}=0.3$.  
Figure~\ref{f.dYe} shows the decrement of the $e^{-}$ abundance, $\Ye(t=0)-\Ye(t) = -\Delta\Ye$ as a function of \carbon\ consumed, $Y_{12}(t=0) - Y_{12}(t) = -\Delta Y_{12}$.  Note that we are plotting the decrement in abundance.  The calculation was started at an initial density and temperature $\rho = 3.0\ee{9}\nsp\grampercc$ and $T = 1.9\ee{8}\nsp\K$.  As $\tH$ shortens, the electron captures onto \sodium[23] ``freeze out'' and $d\Ye/d Y_{12}$ decrease to $\approx 1/6$.  When $-\Delta Y_{12} \gtrsim 0.003$, the abundances of \sodium[23] and \neon[23] have increased sufficiently that they compete with \carbon[12] to consume protons, and thereby halt the neutronization, in agreement with \citet{Piro2007Neutronization-}. 

As noted by \citet{Piro2007Neutronization-}, a one-zone model will
overestimate the heat required to raise the central temperature by a
given amount, and hence overestimate the amount of \carbon\ that must
be consumed during simmering.  We perform a calculation similiar to theirs:
starting with an isothermal white dwarf with a given central density
and with a temperature set by equating heating from the $\carbon
+\carbon$ reaction with neutrino losses, we then raise the central
temperature to $8\ee{8}\nsp\K$, keeping the total white dwarf mass
fixed, and follow an adiabatic temperature gradient until we intersect
the original isotherm at radius $r_{\mathrm{conv}}$, which we then
follow to the stellar surface.  We compute the total stellar energy,
gravitational and thermal, in both cases, and take the difference to
find the heat required to raise the central temperature of the star to
$8\ee{8}\nsp\K$. The temperature of $8\ee{8}\nsp\K$ is chosen as a
fiducial temperature representing the point at which the heating of a fluid element
proceeds faster than the convective timescale
\citep{Woosley2004Carbon-Ignition}.  The evolution of the white dwarf
core is not exactly isobaric: the expanding convective zone heats the
white dwarf. As the entropy of the white dwarf increases, it expands
and reduces the core pressure. For an initial central density
$\rho_{\mathrm{init}} = 3.0\ee{9}\nsp\grampercc$, we find that in raising
the central temperature from $3\ee{8}\nsp\K$ to $8\ee{8}\nsp\K$ the
radius expands by a factor of 1.1 and the central pressure decreases
to 0.59 of its initial value.

For initial central densities $\rho_{\mathrm{init}} =
10^{9}\nsp\grampercc$, $3\ee{9}\nsp\grampercc$, and
$6\ee{9}\nsp\grampercc$, corresponding to white dwarf masses of $1.35\nsp\Msun$,
$1.38\nsp\Msun$, and $1.39\nsp\Msun$, the initial temperatures defined by the onset of thermal instability are $3.8\ee{8}\nsp\K$, $1.9\ee{8}\nsp\K$, and $1.0\ee{8}\nsp\K$, respectively.  The energy required to raise
the central temperature to $8\ee{8}\nsp\K$ is $E_{c} = 2.11$, 4.12, and
5.58\nsp\keV\ per nucleon, respectively.   When the central temperature has reached $8\ee{8}\nsp\K$, the masses of the convective zone for these three cases are $0.69\nsp\Msun$, $1.10\nsp\Msun$, and $1.29\nsp\Msun$, respectively. The spatial extent of the convective zone,  $r_{\mathrm{conv}} \approx 1000\nsp\km$ for the three cases, is in agreement with the findings of \citet{Kuhlen:2006uq}.
We checked our computation of $E_{c}$ by computing $E_{c}' = \int_{\mathrm{conv.}}\,C_{P} \left[ T_{\mathrm{final}}-T_{\mathrm{initial}}\right]\,dM$, as was done by \citet{Piro2007Neutronization-}. Both methods give comparable estimates, but $E_{c}'$ slightly underestimates the change in energy (by $\approx 10\%$), because it does not account for the expansion of the white dwarf. Neglecting the change in \Ye\ as the white dwarf heats introduces a small correction to $E_{c}$: for an adiabatic $1.38\nsp\Msun$ white dwarf with a central temperature of $8\ee{8}\nsp\K$, a reduction in \Ye\ by $1.66\ee{-3}$ reduces $E_{c}$ by only 3.3\%.

If the white dwarf were entirely mixed, raising the central temperature $8\ee{8}\nsp\K$ would require, for the three $\rho_{\mathrm{init}}$ cases here, that $|\Delta Y_{12}| =
1.36\ee{-3}$, $1.66\ee{-3}$, and $1.87\ee{-3}$, respectively.  
Because the changes in composition are only mixed over the convective
zone, the decrement in $Y_{12}$, and hence \Ye, is more pronounced
there.  Using our estimate of $r_{\mathrm{conv}}$, we estimate that
within the convective zone $|\Delta Y_{12}| = 3.0\ee{-3}$, $2.1\ee{-3}$,
and $2.01\ee{-3}$, respectively, for $\rho_{\mathrm{init}}/(10^{9}\nsp\grampercc) = 1.0$, 3.0, and 6.0. Should the radial extent of the convective zone
be smaller than our estimate, for example because of convective Urca
losses \citep{Stein1999The-Role-of-Kin,Stein2006The-Convective-,Lesaffre2005A-two-stream-fo}, the
abundance of \carbon\ will be further reduced in the white dwarf
core. A lower \carbon\ abundance reduces the laminar speed of the flame launched at the end of simmering
\citep{timmes92,Chamulak2007The-Laminar-Fla}.

\begin{figure}[htbp]
\includegraphics[width=86mm]{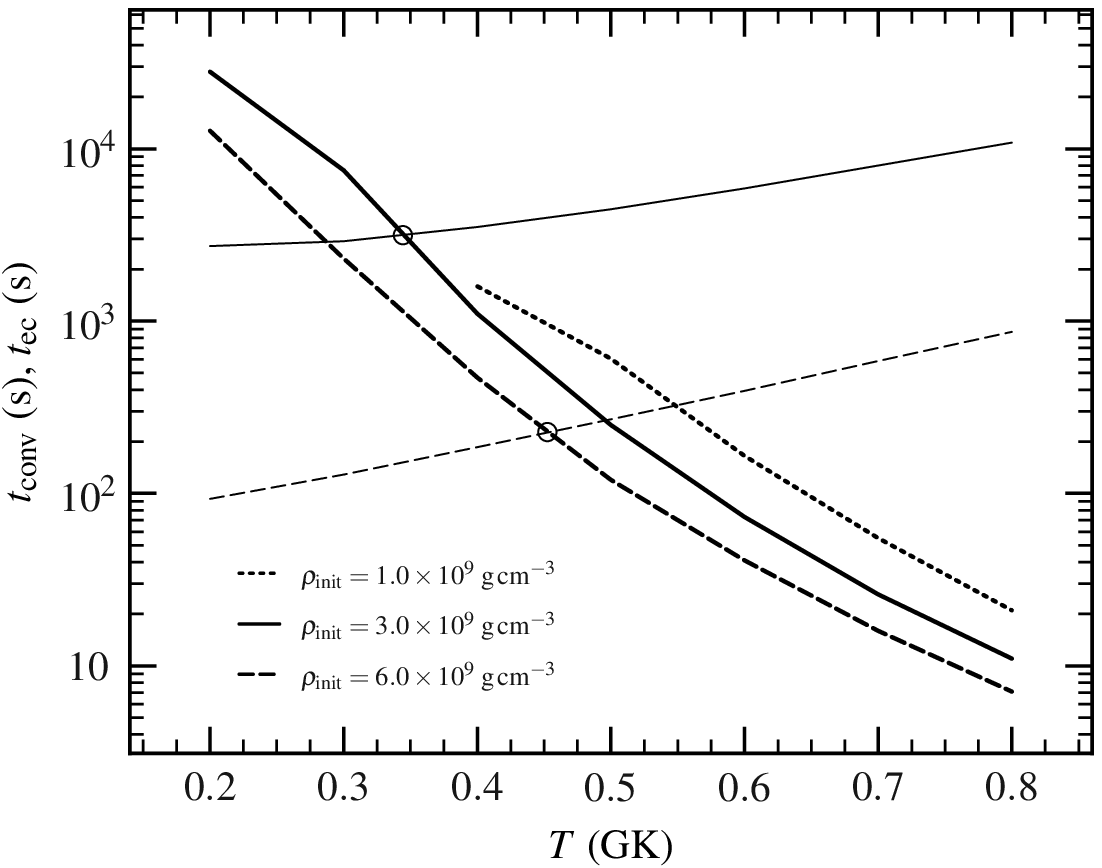}
\caption{Mixing length convective timescales (\emph{thick lines}) and electron capture timescales (for densities above threshold) for \sodium[23] (\emph{thin lines}), as a function of the central temperature during the simmering. Three initial densities are shown: $10^{9}\nsp\grampercc$ (\emph{dotted line}), $3\ee{9}\nsp\grampercc$ (\emph{solid lines}), and $6\ee{9}\nsp\grampercc$ (\emph{dashed lines}). The electron capture timescales increase as the white dwarf heats because the density decreases during simmering.
}
\label{f.time}
\end{figure}

Finally, we estimate at what point the convective mixing timescale becomes shorter than the timescale for electron captures onto \sodium[23]. Using our adiabatic
temperature-gradient white dwarf models, we compute the typical
convective velocity \vrms, and hence a
characteristic turnover time $t_{\mathrm{conv}} =
r_{\mathrm{conv}}/\vrms$, using the mixing length formalism (see the
discussion in \citealt{Woosley2004Carbon-Ignition}) with the total luminosity and evaluating thermodynamical quantities at their central values. Figure~\ref{f.time} shows
$t_{\mathrm{conv}}$ (\emph{thick lines}) for the three cases of $\rho_{\mathrm{init}}$ considered above, as well as the electron capture timescale, $t_{\mathrm{ec}}$ for \sodium[23] (\emph{thin lines}) for
those cases with a density above threshold.  To estimate the effect of the \sodium[23]/\neon[23] pairs on the convective zone, we evaluated the mass fraction of these pairs for the case $\rho_{\mathrm{init}} =3.0\ee{9}\nsp\grampercc$ when the temperature had risen to $T = 3.5\ee{8}\nsp\K$ and $t_{\mathrm{ec}} \approx t_{\mathrm{conv}}$ (Fig.~\ref{f.time}, \emph{solid line}).  At this point the convective core has a mass 0.5\nsp\Msun\, which is comparable to their calculation. We estimate, from the energy required to heat the white dwarf to this point, that the mass fraction of \sodium[23]/\neon[23] pairs in the convective zone will be $X_{23} = 0.004$ at this time. We note that such a large number of Urca pairs will have a dramatic effect on the properties of the convection zone \citep{Lesaffre2005A-two-stream-fo}. As is evident from
Figure~\ref{f.time}, there is a range of temperatures for which
$t_{\mathrm{conv}} < t_{\mathrm{ec}} < \tH$.  In this region, the
effective $d\Ye/dY_{12}$ will depend on the fraction of mass with
densities above the capture threshold and on the effects of Urca
losses on the convective flows, but will still be larger than the
minimum value set by $\nitrogen[13](e^{-},\nu_{e})\carbon[13]$.  Incorporating the effects of the neutrino luminosity from the \sodium[23]/\neon[23] reactions in this regime is numerically challenging \citep{Lesaffre2005A-two-stream-fo,Lesaffre2006The-C-flash-and} and beyond the scope of this work; for now, we just note that this is the primary uncertainty in determining the amount of \carbon\ consumed during the pre-explosive phase, and a better treatment is needed.

\section{Discussion and conclusions}\label{s.discussion}

Using a nuclear reaction network coupled to the equation for
self-heating at constant pressure (eq.~[\ref{eq:T}]--[\ref{eq:Sdot}]),
we have investigated the change in \Ye\ induced by electron captures
on nuclei produced by \carbon\ fusion during the pre-explosion
simmering of the white dwarf. We confirm that there is a maximum \Ye\
at flame ignition \citep{Piro2007Neutronization-}.  We quantified the
role of \neon[22] and other trace nuclides in setting the change in
electron abundance with \carbon\ consumption by using a full reaction
network, and we included the heating from electron captures into an excited state of
\carbon[13].  We gave simple formulae
(eq.~[\ref{e.multiplicity-def}]--[\ref{e.dAbar}]) for the energy
generation rate, the rate of change in electron abundance, and the
rate of change in the mean atomic mass to include the detailed nuclear
physics into large-scale hydrodynamical simulations.  

Our estimates of the maximum \Ye\ at the time of the explosion are roughly similar to those of \citet{Piro2007Neutronization-}.  If we neglect the effect of thermal neutrino losses on the evolution of the white dwarf, then \Ye\ is reduced by $2.7\ee{-4} \textrm{--} 6.3\ee{-4}$ within the convective zone.  This reduction in \Ye\ depends predominantly on the amount of \carbon\ consumed prior to ignition. The electron captures during simmering reduce \Ye\ below the value set by neutron-rich \neon[22] inherited from core He burning by the white dwarf's progenitor star.   Reducing \Ye\ in the explosion
depresses the yield of \nickel[56] and increases the amount of
\iron[54] and \nickel[58] synthesized
\citep{iwamoto.brachwitz.ea:nucleosynthesis, timmes.brown.ea:variations}, even in the absence of further
electron captures onto the Fe-peak isotopes in nuclear statistical
equilibrium (NSE) in the densest portion of the white dwarf.  As a result, any correlation between host system metallicity and white dwarf peak luminosity will be weakened for $Z\lesssim Z_{\odot}$ (for which the reduction in \Ye\ due to captures during simmering is greater than the change due to initial white dwarf composition).

To illustrate how the simmering electron captures affect the
light curve, we reconstruct the comparison made by
\citet{Gallagher2005Chemistry-and-S}, who compiled a set of \SNeIa\
with measured $\Delta m_{15}(B)$, defined as the change in $B$ over 15 days
post-peak, and host galaxies with measured abundances of oxygen to
hydrogen, denoted O/H. We construct an expression for $M_{56}$, the
mass of \nickel[56] produced in the explosion, that depends on $d\Ye/d
Y_{12}$ (Table~\ref{t:Yechange}), $\Delta Y_{12}$, and the host galaxy
composition (we assume that the white dwarf has the same O/H ratio as the
galaxy). The trace nuclide that predominantly sets \Ye\ in the
white dwarf is \neon[22], which traces the aboriginal abundance of CNO
nuclei in the main-sequence star from which the white dwarf
evolved. We therefore recast the linear formula for $M_{56}$
\citep{timmes.brown.ea:variations} in terms of O/H.  For simplicity, we fix the \hydrogen:\helium\ ratio, as well as the ratio of heaver elements to \oxygen, to their solar system
values \citep{Asplund2005The-Solar-Chemi} and neglect corrections from the change in $\mathrm{[O/Fe]}$ with $\mathrm{[Fe/H]}$ \citep{Ramirez2007Oxygen-abundanc} and the increase in $\mathrm{[N/O]}$ with $\mathrm{[O/H]}$ \citep{Liang2006The-Oxygen-Abun}.
With these assumptions, the
mass of \nickel[56] ejected in the explosion is
\begin{equation}\label{eq:M}
M_{56} = M_{56,0} \left[1-72.7\left(\mathrm{\frac{O}{H}}\right) + 58\frac{d\Ye}{d Y_{12}}\Delta Y_{12}\right],
\end{equation}
where $M_{56,0}$ is the total mass of NSE material synthesized at densities where electron captures during the explosion are negligible, $d\Ye/d Y_{12}$ is the change in electron abundance with carbon consumption given in Table~\ref{t:Yechange}, and $-\Delta Y_{12}$ is the total amount of \carbon\ consumed during the pre-explosion convective phase.

\begin{figure}[htpb]
\includegraphics[width=86mm]{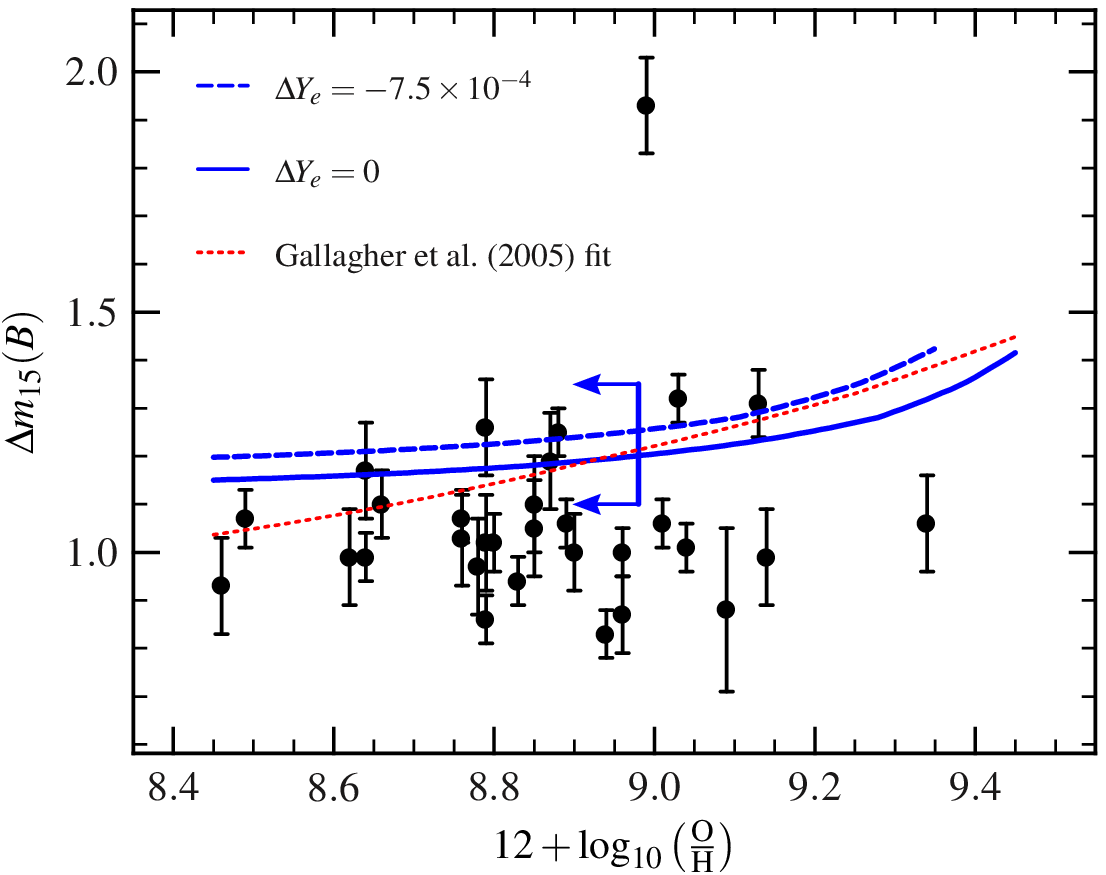}
\caption{Correlation between $12+\log(\mathrm{O/H})$ and $\Delta m_{15}(B)$ induced by the electron abundance. The points in the plot are taken from \citet{Gallagher2005Chemistry-and-S}. We show the 'maximal' case for the reduction in \Ye\ during simmering (\emph{dashed line}), in which $d\Ye/d Y_{12}=0.30$ in the early stages of the burn, which is an upper limit to the derivative.  In this case $\Delta \Ye = -7.5\ee{-4}$.  For comparison, we also show the case in which there is no reduction of \Ye\ during simmering (\emph{solid line}). Note that the relation between $M_{56}$ and $\Delta m_{15}(B)$ is shallower than that used by \citet[\emph{dotted line}]{Gallagher2005Chemistry-and-S}; see the text for an explanation.  To the left of the vertical bar, the decrease in $M_{56}$ due to electron captures during simmering will exceed that due to enrichment by \neon[22], and hence any correlation between O/H and $\Delta m_{15}(B)$ will be masked.}\label{f.dm15}
\end{figure}

Relating this to an observable such as $\Delta m_{15}(B)$ requires an explosion model that follows the radiation transfer. Our interest here is to isolate how $M_{56}$ changes with the CNO abundance of the progenitor when the other parameters, such as the ejecta kinetic energy and the total mass of iron-peak nuclei, are held fixed. This is important, because the relation between light curve width and peak brightness depends on these other  parameters as well \citep[see][for a lucid discussion]{Woosley2006Type-Ia-Superno}.  We use the model M070103 \citep{Woosley2006Type-Ia-Superno}, in which the total mass of iron-peak ejecta is $0.8\nsp\Msun$: of that, the innermost $0.1\nsp\Msun$ is stable iron formed \emph{in situ} from electron captures during the explosion, with the remainder being a mix of radioactive \nickel[56] and stable Fe. Note that this model follows the peak luminosity-light curve width relation \citep{phillips:absolute}, whereas \citet{Mazzali2006The-54Fe58Ni/56} suggest that varying the ratio of \nickel[56] to stable Fe may create dispersion about this relation.

Using the model of \citet{Woosley2006Type-Ia-Superno}, we set $M_{56,0}=0.7\nsp\Msun$, use equation~(\ref{eq:M}) to compute $M_{56}$ as a function of $\mathrm{O/H}$ for different $\Delta Y_{12}$, and interpolate from \citet[Fig.~22]{Woosley2006Type-Ia-Superno} to find $\Delta m_{15}(B)$. Figure~\ref{f.dm15} displays this result.  We plot here a maximal case (\emph{dashed line}) with $d\Ye/d Y_{12} = 0.30$ in the initial part of the simmer, appropriate for the one-zone calculation with electron captures onto \sodium[23] (Fig.~\ref{f.dYe}); for this case $\Delta\Ye = -7.5\ee{-4}$.  For comparison, we also plot a case  (\emph{solid line}) with $\Delta \Ye = 0$ during simmering. This gives a sense of how large the variation in \Ye\ might be. The vertical bar indicates the value of $12+\log_{10}(\mathrm{O/H})$ at which the change in $\Delta m_{15}(B)$ from the \neon[22] abundance equals that from the electron captures during simmering for this maximal case.  To the left of this curve the linear correlation between \neon[22] abundance and $\Delta m_{15}(B)$ will be masked by variations in the simmering of the white dwarf.  For comparison, we also show the data from the compilation of \citet{Gallagher2005Chemistry-and-S} and \citet{hamuy.trager.ea:search} and plot the relation between $12+\log(\mathrm{O/H})$ and $\Delta m_{15}(B)$ used by \citet[\emph{dotted line}]{Gallagher2005Chemistry-and-S}. This trend is much steeper than our finding. The difference is due to how the \nickel[56] mass was varied; whereas the models we use \citep{Woosley2006Type-Ia-Superno} hold the kinetic energy and total mass of iron-peak ejecta fixed, \citet{Gallagher2005Chemistry-and-S} based their peak brightness on delayed detonation models \citep{Hoflich2002Infrared-Spectr} for which a variation in \nickel[56] also produced a change in the relative amounts of iron-peak and intermediate mass-elements, as well as a different explosion kinetic energy.

It is evident from Fig.~\ref{f.dm15} that the scatter in the data points is larger than the expected trend due to progenitor composition, especially at sub-solar metallicities.  Both $\Delta Y_{12}$ and $d\Ye/d Y_{12}$ depend on the central density, which is not obviously correlated with metallicity, and hence the correlation between peak brightness and O/H will be masked by differences in the pre-explosion simmering.  Indeed, if the variation in $\Delta\Ye$ were as large as the two cases we plot in Fig.~\ref{f.dm15}, then variations in \nickel[56] would be determined more by $\Delta Y_{12}$ than by stellar composition for galaxies with sub-solar O/H.  There is a general trend that \SNeIa\ are systematically brighter in galaxies with active star-formation \citep{hamuy.trager.ea:search,Gallagher2005Chemistry-and-S,Sullivan2006Rates-and-prope,Howell2007Predicted-and-o}. \citet{Sullivan2006Rates-and-prope} showed that the \SNeIa\ rate increases linearly with the specific star formation rate, and that \SNeIa\ associated with actively star forming galaxies were intrinsically brighter than those associated with passive galaxies. Although many of these passive galaxies are more massive, and hence more metal-rich \citep{Tremonti2004The-Origin-of-t}, the observed scatter in \SNeIa\ peak brightnesses remains much larger than the expected trend with metallicity (\citealt{Piro2007Neutronization-}; D. A. Howell 2007, private communication). This suggests that the correlation with the chemical abundances of the host galaxy is a secondary effect in setting the peak brightness of \SNeIa.

\acknowledgements 

We thank Tony Piro, Lars Bildsten, Andy Howell, and Craig Wheeler for many useful discussions, Francisco F\"orster and Philipp Podsiadlowski for a detailed comparison of reaction rate networks, Bill Paxton for creating  \code{Tioga} (available at 
\url{http://www.kitp.ucsb.edu/~paxton/tioga.html}), 
and Michele Berry for assistance with Figure~\ref{f.dm15}.  We especially thank Remco Zegers for computing the Gamow-Teller strengths for \nitrogen[13] and the referee for constructive comments. This work was supported by the National Science Foundation, grant AST05-07456, and by the Joint Institute for Nuclear Astrophysics at MSU under NSF-PFC grant PHY02-16783. This research was also supported in part by the NSF under Grant No.~PHY05-51164 to the Kavli Institute for Theoretical Physics. Additional support was provided by the U.~S.~Department of Energy via award KA1401020. Los Alamos National Laboratory is operated by the Los Alamos National Security, LLC for the National Nuclear Security Administration of the U.S. Department of Energy under contract DE-AC52-06NA25396.

\bibliographystyle{apj}
\bibliography{master}

\end{document}